\documentstyle[12pt]{article}

\begin{document}

%\LaTeX{}
{\ }
\bigskip\ \bigskip\ 

\begin{center}
Q1D organic metals - a theoretical determination of the electrical
conductivity\bigskip\ 

Vladan Celebonovic\medskip\ 

Institute of Physics, Pregrevica 118, 11080 Zemun- Beograd, Yugoslavia\smallskip%
\ 

e-mail:celebonovic@exp.phy.bg.ac.yu\bigskip\ \bigskip\ 

\bigskip\ 

received:\bigskip\ \newpage\ 

%2. \bigskip\ % 

\end{center}

Abstract: Using the Hubbard model on a 1D lattice and the ''memory
function'' approach, we have calculated the electrical conductivity of a 1D
system of correlated electrons on a lattice. We have determined the
temperature and frequency profiles of the conductivity. Effects of changes of
the band filling on the conductivity were also briefly discussed. The results
were qualitatively compared with experimental data on the Bechgaard
salts, and the agreement is satisfactory. \bigskip\ \newpage\ 

%\begin{center}

% 3. \bigskip\ 

%\end{center}

Introduction%\smallskip\ 

Studies of correlated electron systems are important for a number of
reasons. In materials of reduced dimensionality, such as the quasi
one-dimensional (Q1D) organic conductors, correlation effects play a highly
important role [1]. High T$_c$ superconductors [2] and the fullerenes [3] are
examples of two and three dimensional materials in which, it is widely
assumed, correlation effects are of crucial importance. 

The aim of this paper is to present results of a calculation of the
contribution of electron-electron scattering to the electrical conductivity
of Q1D organic metals. Theoretical studies of one dimensional correlated
electron systems are interesting in attempts to explain and predict
experimental data on such systems, as well as a possible source of insight in
understanding the mechanism of high T$_c$ superconductivity. The calculations
will refer to the normal (that is, non-superconducting) state of Q1D organic
metals, and they will be qualitatively compared with experimental results on
the Bechgaard salts. The general chemical formula of these materials is
(TMTSF)$_2$X, where (TMTSF)$_2$ stands for
bis-tetramethyl-tetraselenafulvalene, and X is an anyon [4], [5]. 

A few years after the synthesis of the Bechgaard salts, it was shown that
experimental results on their electrical conductivity can not be described
by the standard theory of transport processes in normal metals [7] This
result is the experimental motive for the present calculation. Following
this, and a theoretical prescription by Anderson [8] ,  Q1D organic metals are
usually studied within the Hubbard model. Although this model is one of the
simplest in solid state physics, it is still a subject of intensive study
(see [9]-[13] for some examples).  General reviews of the field of organic
conductors can be found in [14]-[16]. 

All the calculations in this paper were performed by the ''memory function''
method. The main equations needed for the calculation of the conductivity
within this method are briefly presented in the next section, which also
contains a derivation of the expressions for the real and immaginary parts
of the conductivity. Details of the calculation are presented in the third
part, while the fourth section is devoted to a qualitative comparison with
the experimental data. 

The calculations were performed using S. Wolfram's MATHEMATICA software
package, version 1. 2 on a PC 486 with 16 MByte RAM at 133~MHz. \newpage\ 

%\begin{center}
%4. \bigskip\ 
%\end{center}

The method\medskip\ 

In studies of transport phenomena in various kinds of physical systems, one
frequently enounters the problem of having to formulate and solve an
appropriate transport equation. The memory function method gives the
possibility of obtaining the conductivity without having to solve transport
equations, but, instead, by representing the susceptibility in terms of a
suitably chosen memory function. Details of the method are avaliable in the
literature (such as [17], [18] and references given there). 

The calculation of the conductivity is founded on the following two
equations:\\

\begin{equation}
\label{(1)}\chi _{AB}=\ll A;B\gg =-i\int_0^\infty \exp (izt)\langle
[A(t), B(0)]\rangle dt 
\end{equation}
\\and

\begin{equation}
\label{(2)}\sigma (\omega )=i(\omega _P^2/4\pi z)[1-\chi (z)/\chi _0] 
\end{equation}
\medskip\ \\

In eq, (2) $\omega _P^2=4\pi n_ee^2/m_e$ is the square of the plasma
frequency, $n_e$, $e$ and $m_e$ are the electronic number density, charge and
mass , while $\chi _0$ $=n_e/m_e$ is the zero frequency limit of the
dynamical susceptibility $\chi (\omega ). $

The integral in eq. (1) is a general definition of the linear response of an
operator A to a perturbing operator B, and it is an analytic function for all
non-real frequency z [17]. A(t) is the Heisenberg representation of the
operator A. Inserting A=B=[j, H], where j denotes the current operator and H
the Hamiltonian into eq. (1), one gets a definition of the current-current
correlation function.  The chemical potential is a function of the
band-filling n, the inverse temperature $\beta $ and the hopping integral t. 
The form of this function for the 1D Hubbard model on a lattice with lattice
spacing s has recently been determined as [20]:\\\newpage\ 

%\begin{center}
%5. \bigskip\ 

\begin{equation}
\label{(3)}\mu =[(\beta t)^6(ns-1)\mid t\mid ]/[1. 1029+. 1694(\beta
t)^2+. 0654(\beta t)^4] 
\end{equation}
\medskip\ \\
%\end{center}

Taking the lattice constant s=1, and in the special case of half-filling
(n=1), it follows from eq. (3) that $\mu $ = 0, which is in agreement with
standard results [19].  As a first approximation, it will be assumed
throughout this paper that the electrons form a normal Fermi liquid
(however, compare [21] and [22]). 

The Hubbard Hamiltonian has the following form\\

\begin{equation}
\label{(4)}H=-t\sum_{l, \sigma }(c_{l, \sigma }^{+}c_{l+1, \sigma
}+c_{l+1, \sigma }^{+}c_{l, \sigma })+U\sum_jn_{j, \uparrow }n_{j, \downarrow } 
\end{equation}
\medskip\ \\which can be represented as $H=H_0+H_1$, the first term being due
to electronic hopping, and the second one to the inter-electronic on-site
repulsion. \ \\

\ The current operator is given by\\

\begin{center}
\begin{equation}
\label{(5)}j=-it\sum_{l, \sigma }(c_{l, \sigma }^{+}c_{l+1, \sigma
}-c_{l+1, \sigma }^{+}c_{l, \sigma }) 
\end{equation}
\medskip\ \\
\end{center}

The functions $\sigma (z)$ and $\chi (z)$ can be expressed in complex form
as:$\sigma (z)=\sigma _R(z)+i\sigma _I(z)$ and $\chi (z)=\chi _R(z)+i\chi
_I(z). $Assuming that z$=z_1+iz_2$ , and z$_2=\alpha z_1$ with $\alpha \succ
0,  $ it follows from eq. (2) that\\\newpage\ 

%\begin{center}
%6. \bigskip\ 

\begin{equation}
\label{(6)}\sigma _R+i\sigma _I=i\frac{\omega _P^2}{4\pi (1+i\alpha )z_1}[1- 
\frac{\chi _R+i\chi _i}{\chi _0}] 
\end{equation}
%\end{center}

It follows after some obvious algebraic manipulations that the real and
immaginary components of the electrical conductivity are given by the
expressions

\begin{center}
\begin{equation}
\label{(7)}\sigma _R=\frac{\omega _P^2}{4\pi (1+\alpha ^2)z_1}[\alpha (1- 
\frac{\chi _R}{\chi _0})+\frac{\chi _I}{\chi _0}] 
\end{equation}
\end{center}

and

\begin{equation}
\label{(8)}\sigma _I=\frac{\omega _P^2}{4\pi (1+\alpha ^2)z_1}[1-\frac{\chi
_R+\alpha \chi _I}{\chi _0}] 
\end{equation}
\\In the special case $\alpha =0, $these two formulas reduce to

\begin{equation}
\label{(9a)}\sigma _R=\frac{\omega _P^2\chi _I}{4\pi z_1\chi _0} 
\end{equation}
\\and

\begin{equation}
\label{(10)}\sigma _I=\frac{\omega _P^2}{4\pi z_1}(1-\frac{\chi _R}{\chi _0}%
) 
\end{equation}
\\The calculation of the conductivity is in principle straightforward, but
technically complicated. The difficulties arise due to the fact that after
inserting all the ''ingredients'' into eq. (1), one gets an almost intractable
expression. \medskip\ 

The calculations\medskip\ 

In order to calculate the conductivity, one has at first to determine the
current-current correlation function. It can be obtained by inserting $%
A=B=[j, H]$ into eq. (1). \\\newpage\ 

%\begin{center}
%7. \bigskip\ 
%\end{center}

The evaluation of this commutator is simplified to some extent by the
decomposition of the Hubbard Hamiltonian indicated after eq. (4). It can be
shown that $[j, H_0]=0, $ and that\\

\begin{equation}
\label{(11)}A=[j, H]=[j, H_1]=-itU\sum_{l, \sigma }(c_{l, \sigma
}^{+}c_{l+1, \sigma }+c_{l+1, \sigma }^{+}c_{l, \sigma })(\delta
_{l+1, j}-\delta _{l, j})n_{j, -\sigma } 
\end{equation}

where all the symbols have their usul meaning. Transition to k-space can be
performed by the relations of the following general form [6] : 
\begin{equation}
\label{(12)}c_{l, \sigma }^{+}=N^{-1/2}\sum_{k_1}\exp
(ik_1ls)c_{k_1}^{+}, _\sigma 
\end{equation}

The symbol N denotes the number of lattice sites, s is the lattice constant
and L=Ns is the length of the specimen. Using the fact that in the case of
the 1D Hubbard model $\epsilon (k)=-2t\cos (ks), $and introducing the
temporal evolution by relations of the form $c_k(t)=\exp (-i\epsilon
(k)t)c_k $ , one gets the following final expression for $\chi (z)$\\

$\chi (z)=\sum_{p, g, k, q}(32i(1/[(1+\exp (\beta (-\mu -2t\cos (g))))(1+\exp
(\beta (-\mu -2t\cos (k))))]$\smallskip\ 

$-1/[(1+\exp (\beta (-\mu -2t\cos (p))))(1+\exp (\beta (-\mu -2t\cos
(q))))](Ut)^2$\smallskip\ 

$(\alpha z_1+i(z_1+2t(\cos (q)+\cos (p)-\cos (g)-\cos (k))))$\smallskip\ 

$(\cos (p+g)/2)(\cos ((q+k)/2)[\cosh (g-p)-1]/$

%\smallskip\ 

\begin{equation}
\label{(13)}
(N^4((\alpha z_1)^2+(z_1+2t(\cos (q)+\cos (p)-\cos (g)-\cos (k))))^2))
\end{equation}
\medskip\ \ \ \qquad \qquad \qquad \quad 
%(13)

The summations were limited to the first Brioullin zone, and the lattice
constant s has been set equal to 1, which means that L = N. Calculating the
sums in eq. (13) is a non-trivial problem. They have to be evaluated for $%
\alpha \neq 0, $because this condition is built-in into the definition of the
function $\chi (z)$. As the frequency is a real physical quantity, in the
final expression for the dynamical susceptibility the limit $\alpha
\rightarrow 0$ will be imposed. \\\newpage\ 

%\begin{center}
%8. \bigskip\ 
%\end{center}

Performing the summations in eq. (13) by MATHEMATICA, one gets the following
approximate expression for the dynamical susceptibility:\\

$\chi \cong (32i(-1+\cosh (1))[(1+\exp (\beta (-\mu +2t\cos (1-\pi
))))^{-2}- $\smallskip\ 

$(1+\exp (\beta (-\mu -2t)))^{-2}](Ut/N^2)^2\cos {}^2((1-2\pi )/2)[(z_2+$%
\smallskip\ 

$i(z_1+2t(-2-2\cos (1-\pi ))))/(z_2^2+$\smallskip\ 

\begin{equation}
\label{(14)}
(z_1+2t(-2-2\cos (1-\pi )))^2+\ll 2267\gg \medskip\ \qquad \qquad \qquad
\qquad \qquad \qquad \qquad \quad
% \quad (14)
 \end{equation}

\medskip\ 

where $<<2267>>$denotes the number of omitted terms. This equation is
obviously untractable because of its length, and it has to be truncated after
a certain number of terms. Limiting eq. (14) to its first 32 terms, 
multiplying out the products and powers, and expressing the result as a sum
, it follows that the real part of the dynamical susceptibility is \\

%\begin{equation}
%\label{(15)}
$\chi _R\cong [128U^2t^3\cos {}^2((1-2\pi )/2)]/[(1+\exp (\beta (-\mu
-2t)))^2N^4(z_2^2+(z_1+2bt)^2]$
%\end{equation}

\begin{equation}
\label {(15)}
+\ll 527\gg 
\end{equation}

% \qquad \qquad \quad \medskip\ \qquad \qquad \qquad \qquad
%\qquad \qquad \qquad \qquad \qquad \qquad \qquad \qquad \quad 
%(15)$

and $b=2(-2-2\cos (1-\pi )). $

Taking the limit $z_2\rightarrow 0$ (because the frequency is a real
quantity), and limiting eq. (15) to its first 20 terms, it follows finally that
the real component of the dynamical susceptibility can be expressed as
follows:\\

$\chi _R(\omega )=$\ $\sum_iK_i/(\omega +bt)^2+\sum_jL_j\omega /(\omega
+bt)^2\qquad \qquad \qquad \ \qquad \qquad  (16)$

\smallskip\ 

The explicite form of the functions $K_i$and $L_j$ in eq. (16) can be
read-off from eq. (15) developed to any given number of terms. 

The immaginary part of the dynamical susceptibility ( denoted by $\chi
_I)\quad $can be calculated as the following integral [31] :\\\newpage\ 

%\begin{center}
%9. \\
%\end{center}

\bigskip\ 

$\chi _I(\omega _0)=-2(\omega _0/\pi )P\int_0^\infty \chi _R(\omega )d\omega
/(\omega ^2-\omega _0^2)\qquad \qquad \qquad \qquad \qquad 
 (17)$

\smallskip\ 

Limiting the summations in eq. (16) to terms with $i, j\leq 4$ using eq. (17)
and taking the coefficients K and L from eq. (15) , one arrives at the
following expression for $\chi _I$:\\\smallskip\ 

$\chi _I=(2bt/\pi )(Ut/N^2)^2[\omega _0/(\omega _0+2bt)(\omega
_0^2-(2bt)^2)][4. 53316(1+$\smallskip\ 

$\exp (\beta (-\mu -2t)))^{-2}+$24. 6448$(1+\exp (\beta (-\mu +2t\cos (1-\pi
))))^{-2}]+$\smallskip\ 

$(2/\pi )[\omega _0/(\omega _0^2-(2bt)^2)](Ut/N^2)^2$\smallskip\ 

$[42. 49916(1+\exp (\beta (-\mu -2t)))^{-2}+78. 2557(1+$\smallskip\ 

$\exp (\beta (-\mu +2t\cos (1-\pi ))))^{-2}]\qquad \qquad \qquad \qquad
\qquad \qquad \qquad \qquad  \qquad (18)$

\smallskip\ 

Inserting this result into eq. (9), one gets the following expression for the
electrical conductivity of Q1D organic metals:

$\sigma _R(\omega _0)=\omega _P^2\chi _I/4\pi \omega _0\chi _0=$\smallskip\ 

$=(1/2\chi _0)(\omega _P^2/\pi )[\omega
_0^2-(bt)^2]^{-1}(Ut/N^2)^2\{42. 49916(1+\exp (\beta (-\mu -2t)))^{-2}+$%
\smallskip\ 

$78. 2557(1+\exp (\beta (-\mu +2t\cos (1+\pi ))))^{-2}+(bt/(\omega _0+bt))$%
\smallskip\ 

$[4. 53316(1+\exp (\beta (-\mu -2t)))^{-2}+24. 6448\times $

$(1+\exp (\beta (-\mu
+2t\cos (1+\pi ))))^{-2}]\}\qquad\qquad\qquad\qquad\qquad\qquad\qquad \quad (19)$

\medskip\ 

Discussion\smallskip\ 

Equation (19) is the final result of this paper for the real part of the
electrical conductivity of Q1D organic metals. In order to test its
validity, it has to be compared with real experimental data on the Bechgaard
salts. 

This comparison will be performed with respect to three different
experimentally controlable variables:

- the temperature, 

- doping, and

- the frequency. \\\newpage\ 

%\begin{center}
%10. \\
%\end{center}

Model parameters in eq. (19) were chosen as follows: N = 150 ; U = 4 t ;$%
\omega _P=3U;\chi _0=1/3$ and $\omega _{0\geq }. 6U. $ These values were at
first chosen by analogy with high temperature superconductors. In the course
of the calculations it emerged that similar values were used for the
Bechgaard salts by other authors [25]-[27].  The lower limit for $\omega _0$
was determined by imposing the condition $\sigma _R\geq 0$ on eq. (19). 

Changes of the conductivity with the doping can be studied by varying the
chemical potential ,  which, among other factors depends on the band
filling. The case $n=1$ (i. e. , $\mu =0$) corresponds to a half filled band and
, experimentally ,  to a ''clean'' material. Deviations of the filling from the
value $n=1$ describe the doping of th specimen. Positive deviations
correspond to doping by electron donors, while the opposite case (negative
deviations ) describes the doping of a specimen by electron acceptors. 

As a first test. eq. (19) was applied to the case of a half filled
band. Inserting $n=1$ into this expression and developing in $t$ as a small
parameter, one gets that the conductivity is approximately given by $\sigma
_R\cong 10^{-7}(\omega _P/\omega )^2t^4(4. 56+. 3\beta t)$, which is close to
zero for physically acceptable values of the input parameters. This example
can be extended to the level of a general conclusion - that the half filled
1D Hubbard model is an insulator, in line with known results such as [22] or
[29]. Relating to experiments, this implies that weakly conducting phases of
Q1D organic metals can be described by a 1D Hubbard model whose band filling
weakly deviates from 1/2. 

The behaviour of the conductivity as a function of the temperature for
various values of the band filling $n$ was studied by fixing all the
parameters of the system, and then applying eqs. $(3)$ and $(19)$.  The aim of
the calculation was not to fit in detail the experimental data on any
particular Bechgaard salt, but, instead, to reproduce the general trends
observed in resistivity data on these salts.  Examples of experimental curves 
$\rho (T)$ (where $\rho $ denotes the resistivity and T the temperature) are
shown on the following figures. Data on $(TMTSF)_2ReO_4$ and $(TMTSF)_2FSO_3$
come from the author's work [23], while Fig. 3 is taken from [30]. Numbers near
the curves on Fig. 1 indicate values of pressure,  (kilobars), at which the
data were taken. Theoretical curves $\rho (T)$ calculated by expressions (3)
and (19) for various values of the input parameters are shown on figures 4. 
- 10.  Values of the parameters are indicated on the figure captions, and $%
\rho =1$ at T= 116 K. \newpage\ 

%\begin{center}
%11. \\
%\end{center}

A comparison of figures 1 and 2 with figs. 4. -7.  shows that data on $%
(TMTSF)_2ReO_4$ and $(TMTSF)_2FSO_3$ correspond to values of the band
filling between $. 7$and $. 9. $Data on the Bechgaard salts containing the
anions $ClO_4$ and $PF_6$ are best described by the theoretical curve with $%
n=1. 2$. Figures 8 and 9 are, by their form, similar to the form of the
experimental curves $\rho (T)$ for the salts of the $(TMTTF)_2X$ type (fig. 2
). The general conclusion is that the calculations described in this paper
reproduce semi-quantitatively the general behaviour of the experimental
resistivity versus temperature data for the Bechgaard salts. A better
quantitative agreement could be obtained by making an algorithm in which all
of the system parameters could be varied, and the resistivity calculated at
each step.  A comparison of figures 7 and 8 illustrates how a small change of
the system parameters can induce a drastic change in its conductivity, which
a known experimental fact. 

In order to study the behaviour of the conductivity of a Q1D organic metal
as a function of the frequency, eq. (19) can be re-formulated as\medskip\ 

$\sigma _R(\omega _0)=A[\omega _0^2-(bt)^2]^{-1}[Q+btZ(\omega
_0+bt)^{-1}]\qquad \qquad \qquad \qquad \qquad \qquad (20)$

\medskip\ 

The symbols A, Q and Z denote the non-frequency dependent functions in
eq. (19). Developing the last expression in $\omega _0$ up to second order, one
gets \\

$\sigma _R(\omega _0)\cong -A(Q+Z)(bt)^{-2}+AZ(bt)^{-3}\omega
_0-A(2Z-Q)(bt)^{-4}\omega _0^2+. . . . . \qquad (21)$

\medskip\ 

Fitting eq. (21) to the measured frequency profiles of the conductivity, such
as [24] - [27], one could determine the functions A, Q, Z and t. Taking the
limit $\omega _0\rightarrow 0$ in eq. (21), it follows that\medskip\ 

$\sigma _R(\omega _0=0)\cong -A(Q+Z)(bt)^{-2}$ \qquad \qquad \qquad $\qquad
\qquad \qquad \qquad \qquad (22)$

\medskip\ \\which is the approximate static limit of the conductivity. 

Fixing all the parameters, and then varying the band filling, gives the
possibility of investigating the effects of doping on the
resistivity. Examples of changes of resistivity with doping are presented on
figures 11. - 14. Values of input parameters are indicated on each of the
figures. \\\newpage\ 

%\begin{center}
%12. \\
%\end{center}

Clearly, small variations of input parameters lead to large changes of the
conductivity. Note that in all the examples shown on the figures the
resistivity apparently tends towards a ''saturation'' value for band
fillings greater than 1 (that is,  in the case of doping a specimen with
electron donors). \ 

On the other hand,  for some values of the input parameters, doping with
electron acceptors (i. e. , reducing the band filling) leads to a decrease of
the resistivity. This imples that in certain regions of the $\beta -t-\rho $
space, one could achieve increases of conductivity (or, perhaps,  a transition
to superconductivity) by doping the specimen with suitably chosen
impurities. Details of this problem will be discussed elsewhere. \medskip\ 

Conclusions\smallskip\ 

In this paper we have performed a calculation of the electrical conductivity
of Q1D organic metals. The calculation was performed by the ''memory
function'' method, using the 1D Hubbard model on a lattice, and the results
were qualitatively compared with experimental data on the Bechgaard salts. 
The correlation functions were calculated by definition, which is a distinct
advantage over some previous work (such as [28]). Theoretical temperature and
frequency profiles of the conductivity are in good semi-quantitative
agreement with experimental data. Note that the agreement is achieved when
the band filling deviates by about 20\% from 1/2, confirming earlier results
[29].  The calculation was performed within the Fermi liquid picture, which is
justifiable by the energy range we were interested in (compare [26], [30] ). 

Work discussed in this paper will be continued in the future along two
different lines. On the purely calculational side, efforts will be made to
take into account more terms in eq. (14). In improving the physics, a
transition from the Fermi liquid to the Luttinger liquid description will be
made. \newpage\ 

%\begin{center}
%13. \\
%\end{center}

\quad \quad \quad

\quad \quad

\end{document}